\documentclass[10pt]{revtex4}
\pdfoutput=1
\usepackage{amssymb}
\usepackage{latexsym}
\usepackage{epsfig}
\begin{document}

\title{ The origin of Chern-Simons modified gravity from an 11 + 3-dimensional manifold  }

\author{J. A. Helay\"{e}l-Neto  $^{1}$ \footnote{helayel@cbpf.br}, Alireza Sepehri $^{2,3}$\footnote{alireza.sepehri@uk.ac.ir}}
 \affiliation{ $^{1}$ Centro Brasileiro de Pesquisas Físicas, Rua Dr. Xavier Sigaud 150, Urca, Rio de Janeiro, Brazil, CEP 22290-180. \\$^{2}$Faculty of
Physics, Shahid Bahonar University, P.O. Box 76175, Kerman,
Iran.\\$^{3}$ Research Institute for Astronomy and Astrophysics of
Maragha (RIAAM), P.O. Box 55134-441, Maragha, Iran.}

\begin{abstract}
	In this contribution, it is our aim to show that the Chern-Simons terms of modified gravity can be understood as generated by the addition of a 3-dimensional algebraic manifold to an initial 11-dimensional space-time manifold; this builds up an 11+3-dimensional space-time. In this system, firstly, some fields living in the bulk join the fields that live on the 11-dimensional manifold, so that the rank of the gauge fields exceeds the dimension of the algebra; consequently, there emerges an anomaly. To solve this problem, another 11-dimensional manifold is included in the 11 +3-dimensional space-time, and it interacts with the initial manifold by exchanging Chern-Simons fields. This mechanism is able to remove the anomaly. Chern-Simons terms actually produce an extra manifold between the pair of 11-dimensional manifolds of the 11+3-space-time. Summing up over the topology of both the 11-dimensional manifolds and the topology of the exchanged Chern-Simons manifold in the bulk , we conclude that the total topology shrinks to one, which is in agreement with the main idea of the Big Bang theory.


 \end{abstract}
 \date{\today}

\maketitle
\section{Introduction}
 Some authors have recently extended General Relativity and proposed a Chern-Simons modified gravity in which the Einstein-Hilbert action is supplemented by a parity-violating Chern-Simons term, which couples to gravity via a scalar field.  The parity-violating Chern-Simons term is defined as a contraction of the Riemann curvature tensor with its dual and the Chern-Simon scalar field \cite{tt1}. Ever since, a great deal of contributions and discussions on this particular model has appeared in the literature. For example, the authors of Ref. \cite{tt2} have studied the combined effects of the Lorentz-symmetry violating Chern-Simons and Ricci-Cotton actions for the Einstein-Hilbert model in the second order formalism extended by the inclusion of higher-derivative terms, and considered their consequences on the spectrum. In another investigation, the authors have argued about rotating black hole solutions in the (3+1)-dimensional Chern-Simons modified gravity by taking account of perturbations around the Schwarzschild solution \cite{tt3}. They have obtained the zenith-angle dependence of a metric function that corresponds to the frame-dragging effect, by using a constraint equation without choosing the embedding coordinate system. Also,a conserved and symmetric energy-momentum (pseudo-)tensor for Chern-Simons modified gravity has been built up and it has been shown that the model is Lorentz invariant \cite{tt4}.  In another article, the authors have considered the effect of Chern-Simons modified gravity on the quantum phase shift of de Broglie waves in neutron interferometry by applying a unified approach of optical-mechanical analogy in a semiclassical model \cite{tt5}. In a different scenario, the authors have asserted the consistency of the Godel-type solutions within the four-dimensional Chern-Simons modified gravity with the non-dynamical Chern-Simons coefficient, for various shapes of scalar matter and electromagnetic fields \cite{tt6}. Finally, in one of the latest versions of the Chern-Simons gravity, the Chern-Simons scalar fields are treated as dynamical fields possessing their own stress- energy tensor and an evolution equation. This version  has been named Dynamical Chern-Simons Modified Gravity (DCSMG) \cite{tt7,tt8}. Now, a question arises on what this tensor is and and what would be the origin of these Chern-Simons terms. We shall here show that our Universe is a part of an 11-dimensional manifold which is connected with another 11-dimensional manifold by an extra 3-dimensional space. The 11-dimensional manifolds interact with one another via the exchange of Chern-Simons fields which move along the 3-dimensional manifold. 

 Our paper is organized according to the following outline: in Section \ref{o1}, we devote efforts to show that, by adding up a 3-dimensional manifold to eleven- dimensional gravity, there emerges a Chern-Simons modified gravity. Next, in Section \ref{o2}, we shall show that, if the fields obey a special algebra, Chern-Simons modified gravity is shown to be anomaly-free. However, by increasing the rank of the fields, other anomalies show up.  In Section \ref{o3}, we focus on the removal of the anomaly of this type of gravity in a system composed by two 11-dimensional spaces and a Chern-Simons manifold that connects them. In the last Section, we cast a Summary and our Final Considerations.

\section{Chern-Simons modified gravity on an 11+3-dimensional manifold }\label{o1}

We start off by introducing the action of the Dynamical Chern-Simons Modified Gravity \cite{tt7,tt8}:

 \begin{eqnarray}
      && S_{DCSMG} = S_{EH} + S_{CS} + S_{\phi} + S_{mat} \nonumber\\&& S_{EH}= \int d^{4}x \sqrt{-g} R \nonumber\\&& S_{CS}= \int d^{4}x \sqrt{-g}(\frac{1}{2}\epsilon^{\alpha\beta\mu\nu} \phi R_{\alpha\beta\gamma\delta}R^{\gamma\delta}_{\mu\nu} ) \nonumber\\&& S_{\phi}= -\int d^{4}x \sqrt{-g}[g^{\mu\nu}\partial_{\mu}\phi\partial_{\nu}\phi + 2V(\phi)] \label{Os1}
      \end{eqnarray}

where $R$ is the curvature and $\phi$ is the Chern-Simons scalar field.

  Now, we are going to show that Chern-Simons modified gravity can be obtained from a supergravity which lives on an 11+3-dimensional manifold. Actually, we assume that our four-dimensional Universe is a part of an 11-dimensional manifold that interacts with the bulk in an 11+3-dimensional space-time by exchanging Chern-Simons fields.  For this, our departure point is the purely bosonic sector of eleven-dimensional supergravity and wee show that, by adding up a three-dimensional manifold, Chern-Simons terms will appear.

    The bosonic piece of the action for a gravity which lives on an eleven-dimensional manifold is given by \cite{b1,b2}:
          
      \begin{eqnarray}
      && S_{Bosonic-SUGRA} = \frac{1}{\bar{\kappa}^{2}}\int d^{11}x\sqrt{g}\Big(-\frac{1}{2}R-\frac{1}{48}G_{IJKL}G^{IJKL}\Big) + S_{CGG} \nonumber\\&& S_{CGG}=-\frac{\sqrt{2}}{3456\bar{\kappa}^{2}}\int_{M^{11}}d^{11}x \varepsilon^{I_{1}I_{2}...I_{11}}C_{I_{1}I_{2}I_{3}}G_{I_{4}...I_{7}}G_{I_{8}...I_{11}}  \label{s1}
      \end{eqnarray} 
      
      where the curvature ($R$) and $G_{IJKL}$ and $C_{I_{1}I_{2}I_{3}}$, given in terms of the gauge field, $A$, and its field strength, $F$, are cast in what follows below \cite{b2}:

           \begin{eqnarray}
            && G_{IJKL}=-\frac{3}{\sqrt{2}}\frac{\kappa^{2}}{\lambda^{2}}\varepsilon(x^{11})(F_{IJ}F_{KL}-R_{IJ}R_{KL})+... \nonumber\\&& \delta C_{ABC} =-\frac{\kappa^{2}}{6\sqrt{2}\lambda^{2}}\delta (x^{11}) tr( \epsilon_{C} F_{AB}-\epsilon_{C} R_{AB})\nonumber\\&& G_{11ABC}=(\partial_{11}C_{ABC}\pm \text{23 permutations})+\frac{\kappa^{2}}{\sqrt{2}\lambda^{2}}\delta (x^{11})\omega_{ABC}\nonumber\\&& \delta \omega_{ABC}=\partial_{A}(tr \epsilon F_{BC})+ \text{cyclic permutations of A,B,C}\nonumber\\&& F^{IJ}=\partial^{I}A^{J}-\partial^{J}A^{I} \nonumber\\&& R_{IJ}=\partial_{I}\Gamma^{\beta}_{J\beta}-\partial_{J}\Gamma^{\beta}_{I\beta} +\Gamma^{\alpha}_{J\beta}\Gamma^{\beta}_{I\alpha} -\Gamma^{\alpha}_{I\beta}\Gamma^{\beta}_{J\alpha}\nonumber\\&& \Gamma_{IJK}=\partial_{I}g_{JK}+\partial_{K}g_{IJ}-\partial_{J}g_{IK}  \nonumber\\&& G_{IJ}=R_{IJ}-\frac{1}{2}R g_{IJ}\label{s2}
            \end{eqnarray} 
            
      Here,  $\varepsilon(x^{11})$ is 1 for $x^{11}> 0$ and $−1$ for $x^{11}< 0$ and $\delta(x^{11})=\frac{\partial \varepsilon}{\partial x^{11}}$. Both  capitalized Latin (e.g., I, J) and Greek (e.g., $\beta$) indices act on the same manifold and we have only exhibited the free indices I,J,K and the dummy ones  ($\alpha,\beta$). The gauge variation of the CGG-action gives the following result\cite{b2}:

             \begin{eqnarray}
               && \delta S_{CGG}|_{11}=-\frac{\sqrt{2}}{3456\bar{\kappa}^{2}}\int_{M^{11}}d^{11}x \varepsilon^{I_{1}I_{2}...I_{11}}\delta C_{I_{1}I_{2}I_{3}}G_{I_{4}...I_{7}}G_{I_{8}...I_{11}}\approx \nonumber\\&& - \frac{\bar{\kappa}^{4}}{128 \lambda^{6}}\int_{M^{10}}\Sigma_{n=1}^{5}(tr F^{n}-tr R^{n}+tr(F^{n}R^{5-n}))\label{s3}
               \end{eqnarray} 
              
       where, $tr F^{n}=tr(F_{[I_{1}I_{2}}..F_{I_{2n-1}I_{2n}]})=\epsilon^{I_{1}I_{2}..I_{2n-1}I_{2n}}tr(F_{I_{1}I_{2}}..F_{I_{2n-1}I_{2n}})$ and $tr R^{n}=tr(R_{[I_{1}I_{2}}..R_{I_{2n-1}I_{2n}]})=\epsilon^{I_{1}I_{2}..I_{2n-1}I_{2n}}tr(R_{I_{1}I_{2}}..R_{I_{2n-1}I_{2n}})$. These terms above cancel the  anomaly of  ($S_{Bosonic-SUGRA}$) in eleven-dimensional manifold \cite{b2}:
             
       \begin{eqnarray}
                && \delta S_{CGG}|_{11}=-\delta S_{Bosonic-SUGRA}=-\delta S^{anomaly}_{Bosonic-SUGRA}\label{ss3}
                \end{eqnarray}

        Thus, $S_{CGG}$ is necessary for the anomaly cancellation; so, let us now go on and try to find a good rationale for it. Also, we shall answer the question related to the origin of CGG terms in 11-dimensional supergravity. We actually propose a scenario in which the CGG terms appear in the supergravity action in a way that we do not add them up by hand. To this end, we  choose a unified shape for all fields by using the Nambu-Poisson brackets and the properties of string fields ($X$). We define \cite{b4,b9,A1}:

       \begin{eqnarray}
              && I_{J}=\epsilon_{J}I=\epsilon_{J} \frac{\epsilon^{\alpha}_{\alpha J}+\Gamma^{\alpha}_{\alpha J}}{\epsilon^{\alpha}_{\alpha J}+\Gamma^{\alpha}_{\alpha J}}\nonumber\\&& X^{I_{i}}=y^{I_{i}}+A^{I_{i}} + \epsilon^{I_{i}}\phi -\epsilon^{I_{i}J}I_{J}=\nonumber\\&&y^{I_{i}}+A^{I_{i}} + \epsilon^{I_{i}}\phi -\epsilon^{I_{i}J}\epsilon_{J}\frac{\epsilon^{\alpha}_{\alpha J}+\Gamma^{\alpha}_{\alpha J}}{\epsilon^{\alpha}_{\alpha J}+\Gamma^{\alpha}_{\alpha J}}=\nonumber\\&& y^{I_{i}}+A^{I_{i}} + \epsilon^{I_{i}}\phi -\epsilon^{I_{i}J}\Gamma^{\alpha}_{\alpha J}-\epsilon^{I_{i}J}\Sigma_{n=1}^{\infty}(\Gamma^{\alpha}_{\alpha J})^{-n}+...\nonumber\\&& \{ X^{I_{i}},X^{I_{j}} \}=\Sigma_{I_{i},I_{j}}\epsilon^{I'_{i}I'_{j}}\frac{\partial X^{I_{i}}}{\partial y^{I'_{j}}}\frac{\partial X^{I_{j}}}{\partial y^{I'_{j}}}= \nonumber\\&&\Sigma_{I_{i},I_{j}}\epsilon^{I_{i}I'_{i}}\Big(\partial_{I'_{i}}A^{I_{j}}-\partial_{I'_{i}}(\epsilon^{I_{j}I_{k}}\Gamma^{\alpha}_{\alpha I_{k}})+\partial^{I_{i}}\phi\partial_{I_{j}}\phi+..\Big)=F^{I_{i}I_{j}}- R^{I_{i}I_{j}}+\partial^{I_{i}}\phi\partial_{I_{j}}\phi- \frac{1}{2}\epsilon^{I_{i}I_{j}I_{k}I_{m}} \phi R_{I_{i}I_{j}I_{k}I_{m}}+......\label{s4}
              \end{eqnarray} 
            
             where $\phi$ is the Chern-Simons scalar field, $A^{I}$ is the gauge field and $\Gamma$ is related with the curvature (R); I is a unit vector in the direction of the coordinate which can be expanded in terms of derivatives of metric. In fact, the origin of all matter fields and strings is the same and they are equal to the unit vectors ($I_{J}=I\epsilon_{J}$) in addition to some fields ($\phi,A^{I}$) which appear as a fluctuations of space. The latter may emerge by the interaction of strings which breaks the initial symmetric state. Without string interactions, we have a symmetry that could be explained by a unit vector or a matrix. We can first say that, in the static state, all strings are equal to a unit vector or a matrix and, then, these strings interact with one another, so that the symmetry is broken and fields emerge. Using four-dimensional brackets instead of two-dimensional ones, we obtain the shape of the GG-terms in supergravity as functions of strings ($X$):
                 
           \begin{eqnarray}
                    && G^{IJKL}= \{ X^{I},X^{J},X^{K},X^{L} \} =\Sigma_{I'J'K'L'}\epsilon^{I'J'K'L'}\frac{\partial X^{I}}{\partial y^{I'}}\frac{\partial X^{J}}{\partial y^{J'}}\frac{\partial X^{K}}{\partial y^{K'}}\frac{\partial X^{L}}{\partial y^{L'}}\nonumber\\&&\Rightarrow \int d^{11}x\sqrt{g}\Big(G_{IJKL}G^{IJKL}\Big)=\nonumber\\&& \int d^{11}x\sqrt{g}\Big(\Sigma_{IJKL}\epsilon_{I'J'K'L'}\frac{\partial X_{I}}{\partial y_{I'}}\frac{\partial X_{J}}{\partial y_{J'}}\frac{\partial X_{K}}{\partial y_{K'}}\frac{\partial X_{L}}{\partial y_{L'}}\Sigma_{I'J'K'L'}\epsilon^{I'J'K'L'}\frac{\partial X^{I}}{\partial y^{I'}}\frac{\partial X^{J}}{\partial y^{J'}}\frac{\partial X^{K}}{\partial y^{K'}}\frac{\partial X^{L}}{\partial y^{L'}}\Big)  \label{s13}
                    \end{eqnarray} 
                      
                 The equation above helps us to extract the CGG terms from the GG-terms in supergravity. To this end, we must add a three-dimensional manifold (related to a Lie-three-algebra) to eleven-dimensional supergravity by using the properties of strings ($X$) in Nambu-Poisson brackets \cite{A1}:

                     \begin{eqnarray}
                     &&X^{I_{i}}=y^{I_{i}}+A^{I_{i}} -\epsilon^{I_{i}J}\Gamma^{\alpha}_{\alpha J}-\epsilon^{I_{i}J}\Sigma_{n=1}^{\infty}(\Gamma^{\alpha}_{\alpha J})^{-n}\Rightarrow\nonumber\\&& \frac{\partial X^{I_{5}}}{\partial y^{I_{5}}}\approx\delta ( y^{I_{5}})+.. \quad \frac{\partial X^{I_{6}}}{\partial y^{I_{6}}}\approx\delta ( y^{I_{6}})+.. \quad \frac{\partial X^{I_{7}}}{\partial y^{I_{7}}}\approx\delta ( y^{I_{7}})+...\nonumber\\&& \nonumber\\&&\int_{M^{N=3}}\rightarrow\int_{y^{I_{5}}+y^{I_{6}}+y^{I_{7}}}\epsilon^{I'_{5}I'_{6}I'_{7}}\frac{\partial X^{I_{5}}}{\partial y^{I'_{5}}}\frac{\partial X^{I_{6}}}{\partial y^{I'_{6}}}\frac{\partial X^{I_{7}}}{\partial y^{I'_{7}}}=1+.. \label{s14}
                     \end{eqnarray} 
                     
                 where the integration has been carried out over a three-dimensional manifold with coordinates ($y^{I_{5}},y^{I_{6}},y^{I_{7}}$) and, consequently, the integration can be done by using that $\int_{y^{I_{5}}+y^{I_{6}}+y^{I_{7}}}=\int dy^{I_{5}}\int dy^{I_{6}}\int dy^{I_{7}}$). The result above shows that, by ignoring fluctuations of space which yield production of fields, the area of each three-dimensional manifold can shrink to one and the result of the integration over that manifold goes to one. When we add one manifold to the other, the integration will be the product of an integration over each manifold, for the coordinates of the added manifolds increase the elements of integration.  By adding the three-dimensional manifold of equation (\ref{s14}) to the eleven-dimensional manifold of equation (\ref{s13}), we get:

                      \begin{eqnarray}
                      && \int_{M^{N=3}} \times \int_{M^{11}}\sqrt{g}\Big(G_{I_{1}I_{2}I_{3}I_{4}}G^{I_{1}I_{2}I_{3}I_{4}}\Big) = \nonumber\\&& \int_{M^{11}+y^{I_{5}}+y^{I_{6}}+y^{I_{7}}}\sqrt{g}\epsilon^{I'_{5}I'_{6}I'_{7}} G_{I_{1}I_{2}I_{3}I_{4}}G^{I_{1}I_{2}I_{3}I_{4}}\frac{\partial X^{I_{5}}}{\partial y^{I'_{5}}}\frac{\partial X^{I_{6}}}{\partial y^{I'_{6}}}\frac{\partial X^{I_{7}}}{\partial y^{I'_{7}}}= \nonumber\\&&  \int_{M^{11}+M^{N=3}}\sqrt{g}CGG \nonumber\\&&\nonumber\\&&\Rightarrow C^{I_{5}I_{6}I_{7}}= \Sigma_{I'_{5}I'_{6}I'_{7}}\epsilon^{I'_{5}I'_{6}I'_{7}}\frac{\partial X^{I_{5}}}{\partial y^{I'_{5}}}\frac{\partial X^{I_{6}}}{\partial y^{I'_{6}}}\frac{\partial X^{I_{7}}}{\partial y^{I'_{7}}}\label{s15}
                      \end{eqnarray} 
                      
                  This equation present three results we should comment on : 1. CGG terms may appear in the action of supergravity by adding a three-dimensional manifold, related to the Lie-three-algebra added to eleven-dimensinal supergravity. 2. 11-dimensional manifold + three-Lie-algebra = 14-dimensional supergravity. 3. The shape of the C-terms is now clear in terms of the string fields, ($X^{i}$).
                  
                  Substituting equations ( \ref{s4}, \ref{s13} and \ref{s14}) into equation (\ref{s15}) yields:
                                                                                                                          \begin{eqnarray}
      &&  \int_{M^{11}+M^{N=3}}\sqrt{g}CGG = \nonumber\\&& \int_{M^{11}+M^{N=3}}\sqrt{g}\epsilon_{I_{1}I_{2}I_{3}I_{4}I'_{1}I'_{2}I'_{3}I'_{4}I_{5}I_{6}I_{7}}\epsilon^{\tilde{I}_{5}\tilde{I}_{6}\tilde{I}_{7}}(\frac{\partial X^{I_{5}}}{\partial y^{\tilde{I}_{5}}}\frac{\partial X^{I_{6}}}{\partial y^{\tilde{I}_{6}}}\frac{\partial X^{I_{7}}}{\partial y^{\tilde{I}_{7}}}) G^{I_{1}I_{2}I_{3}I_{4}}G^{I'_{1}I'_{2}I'_{3}I'_{4}}= \nonumber\\&&\int_{M^{11}+M^{N=3}}\sqrt{g}\epsilon_{I_{1}I_{2}I_{3}I_{4}I'_{1}I'_{2}I'_{3}I'_{4}I_{5}I_{6}I_{7}}\epsilon^{\tilde{I}_{5}\tilde{I}_{6}\tilde{I}_{7}}(\frac{\partial X^{I_{5}}}{\partial y^{\tilde{I}_{5}}}\frac{\partial X^{I_{6}}}{\partial y^{\tilde{I}_{6}}}\frac{\partial X^{I_{7}}}{\partial y^{\tilde{I}_{7}}}) \times \nonumber\\&&(\epsilon^{\tilde{I}_{1}\tilde{I}_{2}\tilde{I}_{3}\tilde{I}_{4}}\frac{\partial X^{I_{1}}}{\partial y^{\tilde{I}_{1}}}\frac{\partial X^{I_{2}}}{\partial y^{\tilde{I}_{2}}}\frac{\partial X^{I_{3}}}{\partial y^{\tilde{I}_{3}}}\frac{\partial X^{I_{4}}}{\partial y^{\tilde{I}_{4}}})(\epsilon^{\tilde{I}'_{1}\tilde{I}'_{2}\tilde{I}'_{3}\tilde{I}'_{4}}\frac{\partial X^{I'_{1}}}{\partial y^{\tilde{I}'_{1}}}\frac{\partial X^{I'_{2}}}{\partial y^{\tilde{I}'_{2}}}\frac{\partial X^{I'_{3}}}{\partial y^{\tilde{I}'_{3}}}\frac{\partial X^{I'_{4}}}{\partial y^{\tilde{I}'_{4}}})=\nonumber\\&& \int_{M^{11}+M^{N=3}}\sqrt{g}\epsilon_{I_{1}I_{2}I_{3}I_{4}I'_{1}I'_{2}I'_{3}I'_{4}I_{5}I_{6}I_{7}}\epsilon^{\tilde{I}_{5}\tilde{I}_{6}\tilde{I}_{7}}(\frac{\partial X^{I_{5}}}{\partial y^{\tilde{I}_{5}}}\frac{\partial X^{I_{6}}}{\partial y^{\tilde{I}_{6}}}\frac{\partial X^{I_{7}}}{\partial y^{\tilde{I}_{7}}}) \times \nonumber\\&&(\epsilon^{\tilde{I}_{1}\tilde{I}_{2}\tilde{I}_{3}\tilde{I}_{4}}\frac{\partial X^{I_{1}}}{\partial y^{\tilde{I}_{1}}}\frac{\partial X^{I_{2}}}{\partial y^{\tilde{I}_{2}}}\frac{\partial X^{I_{3}}}{\partial y^{\tilde{I}_{3}}}\frac{\partial X^{I_{4}}}{\partial y^{\tilde{I}_{4}}})(\epsilon^{\tilde{I}'_{1}\tilde{I}'_{2}\tilde{I}'_{3}\tilde{I}'_{4}}\frac{\partial X^{I'_{1}}}{\partial y^{\tilde{I}'_{1}}}\frac{\partial X^{I'_{2}}}{\partial y^{\tilde{I}'_{2}}}\frac{\partial X^{I'_{3}}}{\partial y^{\tilde{I}'_{3}}}\frac{\partial X^{I'_{4}}}{\partial y^{\tilde{I}'_{4}}})=\nonumber\\&& \int_{M^{11}+M^{N=3}}\sqrt{g}\epsilon_{I_{1}I_{2}I_{3}I_{4}I'_{1}I'_{2}I'_{3}I'_{4}I_{5}I_{6}I_{7}}(\epsilon^{\tilde{I}_{4}\tilde{I}_{5}\tilde{I}_{7}}\frac{\partial X^{I_{4}}}{\partial y^{\tilde{I}_{4}}}\frac{\partial X^{I_{5}}}{\partial y^{\tilde{I}_{5}}}\frac{\partial X^{I_{7}}}{\partial y^{\tilde{I}_{7}}})(\epsilon^{\tilde{I}'_{4}\tilde{I}_{6}}\frac{\partial X^{I'_{4}}}{\partial y^{\tilde{I}'_{4}}}\frac{\partial X^{I_{6}}}{\partial y^{\tilde{I}_{6}}}) \times \nonumber\\&&(\epsilon^{\tilde{I}_{1}\tilde{I}_{2}}\frac{\partial X^{I_{1}}}{\partial y^{\tilde{I}_{1}}}\frac{\partial X^{I_{2}}}{\partial y^{\tilde{I}_{2}}})(\epsilon^{\tilde{I}'_{1}\tilde{I}'_{2}}\frac{\partial X^{I'_{1}}}{\partial y^{\tilde{I}'_{1}}}\frac{\partial X^{I'_{2}}}{\partial y^{\tilde{I}'_{2}}})(\epsilon^{\tilde{I}_{3}\tilde{I}'_{3}}\frac{\partial X^{I_{3}}}{\partial \tilde{I}^{I_{3}}}\frac{\partial X^{I'_{3}}}{\partial y^{\tilde{I}'_{3}}})= \nonumber\\&& \int_{M^{11}+M^{N=3}}\sqrt{g}\Big(  \frac{1}{2}\epsilon^{I_{i}I_{j}I_{k}I_{m}} \phi R_{I_{k}I_{m}I_{l}I_{n}}R^{I_{l}I_{n}}_{I_{i}I_{j}}- \partial_{I_{i}}\phi\partial^{I_{j}}\phi \Big) - \frac{1}{2} \int_{M^{11}+M^{N=3}}\sqrt{g}\Big(\phi \epsilon^{I_{i}I_{j}}_{I_{k}I_{m}}F^{I_{k}I_{m}}F_{I_{i}I_{j}}\Big)+...\label{Is17}
                                                                                                                                                                \end{eqnarray} 
                                                                                                            In the equation above, the first integration is in agreement with previous predictions of Chern-Simons gravity in \cite{tt7,tt8} and can be reduced to the four-dimensional Chern-Simons modified gravity of equation (\ref{Os1}). Also, the second integration is related to the interaction of gauge fields with Chern-Simons fields. Thus, this model not only produces the Chern-Simons modified gravity, but also exhibits some modifications to it. Still, these results show that our Universe is a part of one-eleven dimensional manifold which interacts with a bulk in a 14-dimensional space-time by exchanging Chern-Simons scalars. 
                                                                                                            
                                                                                                                 \section{Anomalies in Chern-Simons modified gravity  }\label{o2}
                                                                                                                 In this Section, we shall consider various anomalies which may be induced in Chern-Simons modified gravity. Although we expect that terms in the gauge variation of the Chern-Simons action removes the anomaly in eleven-dimensional supergravity, we will observe that some extra anomalies are produced by the Chern-Simons field.  It is our goal to show that these anomalies depend on the algebra and thus, by choosing a suitable algebra  in this model, all anomalies can be removed. To obtain the anomalies of the Chern-Simons theory, we should re-obtain the gauge variation of the CGG-action in equation (\ref{s3}) in terms of field-strengths and curvatures. To this end, by using equation (\ref{s14} and \ref{s15}), we can work out the gauge variation of C \cite{A1}:

                             \begin{eqnarray}
                             &&  X^{I_{i}}=y^{I_{i}}+A^{I_{i}} -\epsilon^{I_{i}J}\Gamma^{\alpha}_{\alpha J}-\epsilon^{I_{i}J}\Sigma_{n=1}^{\infty}(\Gamma^{\alpha}_{\alpha J})^{-n}\Rightarrow \frac{\partial \delta_{A} X^{I}}{\partial y^{I}}=\delta ( y^{I}) \nonumber\\&&\nonumber\\&&\Rightarrow \int_{M^{N=3}+M^{11}}\delta_{A} C^{I_{5}I_{6}I_{7}}=\int_{M^{N=3}+M^{11}} \Sigma_{I'_{5}I'_{6}I'_{7}}\epsilon^{I'_{5}I'_{6}I'_{7}}\delta_{A}(\frac{\partial X^{I_{5}}}{\partial y^{I'_{5}}}\frac{\partial X^{I_{6}}}{\partial y^{I'_{6}}}\frac{\partial X^{I_{7}}}{\partial y^{I'_{7}}})=\nonumber\\&& \int_{M^{N=3}+M^{10}}\Sigma_{I'_{5}I'_{6}}\epsilon^{I'_{5}I'_{6}}(\frac{\partial X^{I_{5}}}{\partial y^{I'_{5}}}\frac{\partial X^{I_{6}}}{\partial y^{I'_{6}}})=\nonumber\\&&\int_{M^{N=3}+M^{10}}(F^{I_{i}I_{j}}- R^{I_{i}I_{j}}+\partial^{I_{i}}\phi\partial_{I_{j}}\phi - \frac{1}{2}\epsilon^{I_{i}I_{j}I_{k}I_{m}} \phi R_{I_{i}I_{j}I_{k}I_{m}}+...)\label{s16}
                             \end{eqnarray}

                                  Using the equation above and the equation (\ref{s13}), we get the gauge variation of the CGG action given in equation (\ref{s15}):
                                 
                                 \begin{eqnarray}
                                      &&  \delta\int_{M^{11}+M^{N=3}}\sqrt{g}CGG = \nonumber\\&& \delta\int_{M^{11}+M^{N=3}}\sqrt{g}\epsilon_{I_{1}I_{2}I_{3}I_{4}I'_{1}I'_{2}I'_{3}I'_{4}I_{5}I_{6}I_{7}}\epsilon^{\tilde{I}_{5}\tilde{I}_{6}\tilde{I}_{7}}(\frac{\partial X^{I_{5}}}{\partial y^{\tilde{I}_{5}}}\frac{\partial X^{I_{6}}}{\partial y^{\tilde{I}_{6}}}\frac{\partial X^{I_{7}}}{\partial y^{\tilde{I}_{7}}}) G^{I_{1}I_{2}I_{3}I_{4}}G^{I'_{1}I'_{2}I'_{3}I'_{4}}= \nonumber\\&&\delta\int_{M^{11}+M^{N=3}}\sqrt{g}\epsilon_{I_{1}I_{2}I_{3}I_{4}I'_{1}I'_{2}I'_{3}I'_{4}I_{5}I_{6}I_{7}}\epsilon^{\tilde{I}_{5}\tilde{I}_{6}\tilde{I}_{7}}(\frac{\partial X^{I_{5}}}{\partial y^{\tilde{I}_{5}}}\frac{\partial X^{I_{6}}}{\partial y^{\tilde{I}_{6}}}\frac{\partial X^{I_{7}}}{\partial y^{\tilde{I}_{7}}}) \times \nonumber\\&&(\epsilon^{\tilde{I}_{1}\tilde{I}_{2}\tilde{I}_{3}\tilde{I}_{4}}\frac{\partial X^{I_{1}}}{\partial y^{\tilde{I}_{1}}}\frac{\partial X^{I_{2}}}{\partial y^{\tilde{I}_{2}}}\frac{\partial X^{I_{3}}}{\partial y^{\tilde{I}_{3}}}\frac{\partial X^{I_{4}}}{\partial y^{\tilde{I}_{4}}})(\epsilon^{\tilde{I}'_{1}\tilde{I}'_{2}\tilde{I}'_{3}\tilde{I}'_{4}}\frac{\partial X^{I'_{1}}}{\partial y^{\tilde{I}'_{1}}}\frac{\partial X^{I'_{2}}}{\partial y^{\tilde{I}'_{2}}}\frac{\partial X^{I'_{3}}}{\partial y^{\tilde{I}'_{3}}}\frac{\partial X^{I'_{4}}}{\partial y^{\tilde{I}'_{4}}})=\nonumber\\&& \int_{M^{10}+M^{N=3}}\sqrt{g}\epsilon_{I_{1}I_{2}I_{3}I_{4}I'_{1}I'_{2}I'_{3}I'_{4}I_{5}I_{6}}\epsilon^{\tilde{I}_{5}\tilde{I}_{6}}(\frac{\partial X^{I_{5}}}{\partial y^{\tilde{I}_{5}}}\frac{\partial X^{I_{6}}}{\partial y^{\tilde{I}_{6}}}) \times \nonumber\\&&(\epsilon^{\tilde{I}_{1}\tilde{I}_{2}\tilde{I}_{3}\tilde{I}_{4}}\frac{\partial X^{I_{1}}}{\partial y^{\tilde{I}_{1}}}\frac{\partial X^{I_{2}}}{\partial y^{\tilde{I}_{2}}}\frac{\partial X^{I_{3}}}{\partial y^{\tilde{I}_{3}}}\frac{\partial X^{I_{4}}}{\partial y^{\tilde{I}_{4}}})(\epsilon^{\tilde{I}'_{1}\tilde{I}'_{2}\tilde{I}'_{3}\tilde{I}'_{4}}\frac{\partial X^{I'_{1}}}{\partial y^{\tilde{I}'_{1}}}\frac{\partial X^{I'_{2}}}{\partial y^{\tilde{I}'_{2}}}\frac{\partial X^{I'_{3}}}{\partial y^{\tilde{I}'_{3}}}\frac{\partial X^{I'_{4}}}{\partial y^{\tilde{I}'_{4}}})=\nonumber\\&& \int_{M^{10}+M^{N=3}}\sqrt{g}\epsilon_{I_{1}I_{2}I_{3}I_{4}I'_{1}I'_{2}I'_{3}I'_{4}I_{5}I_{6}}(\epsilon^{\tilde{I}_{4}\tilde{I}_{5}}\frac{\partial X^{I_{4}}}{\partial y^{\tilde{I}_{4}}}\frac{\partial X^{I_{5}}}{\partial y^{\tilde{I}_{5}}})(\epsilon^{\tilde{I}'_{4}\tilde{I}_{6}}\frac{\partial X^{I'_{4}}}{\partial y^{\tilde{I}'_{4}}}\frac{\partial X^{I_{6}}}{\partial y^{\tilde{I}_{6}}}) \times \nonumber\\&&(\epsilon^{\tilde{I}_{1}\tilde{I}_{2}}\frac{\partial X^{I_{1}}}{\partial y^{\tilde{I}_{1}}}\frac{\partial X^{I_{2}}}{\partial y^{\tilde{I}_{2}}})(\epsilon^{\tilde{I}'_{1}\tilde{I}'_{2}}\frac{\partial X^{I'_{1}}}{\partial y^{\tilde{I}'_{1}}}\frac{\partial X^{I'_{2}}}{\partial y^{\tilde{I}'_{2}}})(\epsilon^{\tilde{I}_{3}\tilde{I}'_{3}}\frac{\partial X^{I_{3}}}{\partial \tilde{I}^{I_{3}}}\frac{\partial X^{I'_{3}}}{\partial y^{\tilde{I}'_{3}}})= \nonumber\\&& \int_{M^{10}+M^{N=3}}\sqrt{g}\Sigma_{n=1}^{5}\Big(tr F^{n}-tr R^{n}+tr(F^{n}R^{5-n})\Big) + \nonumber\\&& \int_{M^{10}+M^{N=3}}\sqrt{g}\Big(\Sigma_{n=1}^{5}(tr( F^{n}(\partial^{I_{i}}\phi\partial_{I_{j}}\phi)^{5-n}) + tr( F^{n}(\epsilon^{I_{i}I_{j}I_{k}I_{m}}  R_{I_{i}I_{j}I_{k}I_{m}}\phi)^{5-n}) \Big) - \nonumber\\&& \int_{M^{10}+M^{N=3}}\sqrt{g}\Big(\Sigma_{n=1}^{5}(tr( R^{n}(\partial^{I_{i}}\phi\partial_{I_{j}}\phi)^{5-n}) + tr( R^{n}(\epsilon^{I_{i}I_{j}I_{k}I_{m}}  R_{I_{i}I_{j}I_{k}I_{m}}\phi)^{5-n}) \Big)+...\label{s17}
                                      \end{eqnarray} 
                                                                                                              The first line of this equation removes the anomaly on the 11-dimensional manifold of equation (\ref{s3}); however, the second and third lines show that extra anomalies can emerge due to the Chern-Simons fields. We can show that, if we choose a suitable algebra for the 11-dimensional manifold, all anomalies can be swept awat. We can extend our discussion to a D-dimensional manifold with a Lie-N-algebra. In fact, we wish to obtain a method that makes all supergravities, with arbitrary dimension, anomaly-free. To this end, we make use of the properties of Nambu-Poisson brackets and  strings ($X$) in equation (\ref{s4}) to obain a unified definition for different terms in supergravity, and rewrite the action (\ref{s3}) as follows:
                                                                                                                
                                                                                                                 \begin{eqnarray}
                                                                                                                    && \delta S_{CGG}=-\frac{\bar{\kappa}^{4}}{128 \lambda^{6}}\int_{M^{10}}\epsilon^{I_{1}I_{2}..I_{10}} \{ X^{I_{1}},X^{I_{2}} \}\{ X^{I_{3}},X^{I_{4}} \}\{ X^{I_{5}},X^{I_{6}} \}\{ X^{I_{7}},X^{I_{8}} \}\{ X^{I_{9}},X^{I_{10}} \}\label{s5}
                                                                                                                    \end{eqnarray}                                                                                          
 
  In the equation above, we only used the Lie-two-algebra with two-dimensional bracket; however, it is not clear that this algebra be true. In fact, for M-theory, Lie-three-algebra with three-dimensional bracket \cite{b4,b9} is more suitable. To obtain the exact form of the Lie-algebra which is suitable for  D-dimensional space-time, we shall generalize the dimension of space-time from eleven to D and the algebra from two to N and use the following Nambu-Poisson brackets \cite{b9}: 
                                                                                                                   \begin{eqnarray}
                                                                                                                         && \int_{M^{D}}\{ X^{I_{i}},X^{I_{j}} \}... \rightarrow \int_{M^{N+D}}\epsilon^{I_{i}I_{j}}_{J_{1}J_{2}...J_{N}} \{ X^{J_{1}},X^{J_{2}},...X^{J_{N}} \}... \label{s6}               \end{eqnarray} 
                                                                                                            In this equation, we have added a new manifold, related to the algebra, to the world manifold. In fact, we have to regard both algebraic ($M^{N}$) and space-time ($M^{D}$) manifolds to achieve the exact results. For the N-dimensional algebra, we introduce the following fields:
                                                                                                           
 \begin{eqnarray}
            &&  X^{J_{N}}\rightarrow y^{J_{N}}+\epsilon^{J_{N}}_{J_{1},J_{2},...J_{N-1}} A^{J_{1},J_{2},...J_{N-1}}-\epsilon^{J_{N}}_{J_{1},J_{2},...J_{N-1}} \partial^{J_{4}}..\partial^{J_{N-1}}\Gamma^{J_{1},J_{2},J_{3}}\nonumber\\&& F^{J_{1}...J_{N}}=\epsilon^{J_{N}}_{J_{1},J_{2},...J_{N-1}} \partial_{J_{N}} A^{J_{1},J_{2},...J_{N-1}}\nonumber\\&& \partial^{J_{5}}...\partial^{J_{N}}R^{J_{1}...J_{4}}=\epsilon^{J_{N}}_{J_{1},J_{2},...J_{N-1}} \partial_{J_{N}} \partial^{J_{4}}...\partial^{J_{N-1}}\Gamma^{J_{1}J_{2}J_{3}}+..\label{sv6}
                       \end{eqnarray}
                       
  where

                 \begin{eqnarray}
                  && \epsilon^{I_{i}I_{j}}_{J_{1}J_{2}...J_{N}}=\epsilon^{I_{i}I_{j}}\epsilon_{J_{1}J_{2}...J_{N}}\nonumber\\&& \epsilon^{I_{i}}_{J_{1}J_{2}...J_{N}}=\delta^{I_{i}}_{[J_{1}J_{2}...J_{N}]}\nonumber\\&&\delta_{[J_{1}J_{2}...J_{N}]}=\delta_{J_{1}J_{2}...J_{N}}-\delta_{J_{2}J_{1}...J_{N}}+....\label{sjlk8}
                             \end{eqnarray}

       Here, $\delta$ is  the generalized Kronecker delta. With  definitions in equation (\ref{sv6}), we can obtain the explicit form of the N-dimensional Nambu-Poisson brackets in terms of fields:

                \begin{eqnarray}
                 &&  \int_{M^{N}+M^{D}}\{ X^{J_{1}},X^{J_{2}},...X^{J_{N}} \}=\int_{M^{N}}\Sigma_{J_{1},J_{2},...J_{N}}\epsilon^{J_{1},J_{2},...J_{N}}\frac{\partial X^{J_{1}}}{\partial y^{J_{1}}}\frac{\partial X^{J_{2}}}{\partial y^{J_{2}}}.....\frac{\partial X^{J_{N}}}{\partial y^{J_{N}}}\approx \nonumber\\&&\int_{M^{N}+M^{D}}(F^{J_{1}...J_{N}}-\partial^{J_{5}}...\partial^{J_{N}}R^{J_{1}...J_{4}}) \label{svv6}
                            \end{eqnarray} 
                                                                                                             Substituting equations (\ref{s6}, \ref{sv6} and \ref{svv6} ) in (\ref{s5}), which is another form of (\ref{s3}), and replacing 11-dimensional manifold with D dimensional manifold, we obtain:
                                                                                                                 
                                                                                                                   \begin{eqnarray}
                                                                                                                      && \delta S_{CGG}|_{D+1}=-Z\int_{M^{D}}\epsilon_{I_{1}I_{2}..I_{D}} \{ X^{I_{1}},X^{I_{2}} \}\{ X^{I_{3}},X^{I_{4}} \}...\{ X^{I_{D-1}},X^{I_{D}} \}=\nonumber\\&& -Z\int_{M^{D+N}}\epsilon_{I_{1}I_{2}..I_{D}}\epsilon^{I_{1}I_{2}}_{J^{1}_{1}J^{1}_{2}...J^{1}_{N}} \{ X^{J^{1}_{1}},X^{J^{1}_{2}},...X^{J^{1}_{N}} \} \epsilon^{I_{3}I_{4}}_{J^{2}_{1}J^{2}_{2}...J^{2}_{N}} \{ X^{J^{2}_{1}},X^{J^{2}_{2}},...X^{J^{2}_{N}} \}...\epsilon^{I_{D-1}I_{D}}_{J^{D/2}_{1}J^{D/2}_{2}...J^{D/2}_{N}} \{ X^{J^{D/2}_{1}},X^{J^{D/2}_{2}},...X^{J^{D/2}_{N}} \}\nonumber\\&& =-Z\int_{M^{D+N}}\epsilon_{I_{1}I_{2}..I_{D}}\epsilon^{I_{1}I_{2}}_{J^{1}_{1}J^{1}_{2}...J^{1}_{N}}\epsilon^{I_{3}I_{4}}_{J^{2}_{1}J^{2}_{2}...J^{2}_{N}}...\epsilon^{I_{D-1}I_{D}}_{J^{D/2}_{1}J^{D/2}_{2}...J^{D/2}_{N}} (F^{J^{1}_{1}...J^{1}_{N}}-\partial^{J^{1}_{5}}...\partial^{J^{1}_{N}}R^{J^{1}_{1}...J^{1}_{4}}) \times \nonumber\\&& (F^{J^{2}_{1}...J^{2}_{N}}-\partial^{J^{2}_{5}}...\partial^{J^{2}_{N}}R^{J^{2}_{1}...J^{2}_{4}})... (F^{J^{1}_{D/2}...J^{D/2}_{N}}-\partial^{J^{D/2}_{5}}...\partial^{J^{D/2}_{N}}R^{J^{D/2}_{1}...J^{D/2}_{4}})\label{s9}
                                                                                                                      \end{eqnarray}
                                                                                                               where $Z$ is a constant related to the algebra. This equation shows that the gauge variation of the action depends on the rank-N field-strength. The action above is not actually directly zero, and there emerges an anomaly.  Now, we use properties of $\epsilon$ and rewrite equation (\ref{s9}) as below:
                                                                                                                            \begin{eqnarray}                                                                                              && \delta S_{CGG}|_{D+1}=\nonumber\\&&-Z\int_{M^{D+N}}W(D,N)\epsilon^{\chi_{1}\chi_{2}..\chi_{D/2}} (F^{\chi_{1}}-\partial^{J^{1}_{5}}...\partial^{\chi_{1}-4}R^{J^{1}_{1}...J^{1}_{4}}) \times \nonumber\\&& (F^{\chi_{2}}-\partial^{J^{2}_{5}}...\partial^{\chi_{2}-4}R^{J^{2}_{1}...J^{2}_{4}})... (F^{\chi_{D/2}}-\partial^{J^{D/2}_{5}}...\partial^{\chi_{D/2}-4}R^{J^{D/2}_{1}...J^{D/2}_{4}})\label{s10}
                                                                                                                               \end{eqnarray}
                                                                                                                In equation (\ref{s10}), $\chi$, $\epsilon^{\chi_{1}\chi_{2}..\chi_{D/2}}$ and $W(D,N)$ can be obtained as:
                                                                                                                                                                                                                            \begin{eqnarray}                                                                                                            && \chi_{i}=J^{i}_{1}...J^{i}_{N}\nonumber\\&&\nonumber\\&&\epsilon^{I_{1}I_{2}..I_{D}}\epsilon^{I_{1}I_{2}}_{J^{1}_{1}J^{1}_{2}...J^{1}_{N}}\epsilon^{I_{3}I_{4}}_{J^{2}_{1}J^{2}_{2}...J^{2}_{N}}...\epsilon^{I_{D-1}I_{D}}_{J^{D/2}_{1}J^{D/2}_{2}...J^{D/2}_{N}}=\nonumber\\&&\epsilon^{I_{1}I_{2}..I_{D}}\epsilon^{I_{1}I_{2}}\epsilon_{J^{1}_{1}J^{1}_{2}...J^{1}_{N}}\epsilon^{I_{3}I_{4}}\epsilon_{J^{2}_{1}J^{2}_{2}...J^{2}_{N}}...\epsilon^{I_{D-1}I_{D}}\epsilon_{J^{D/2}_{1}J^{D/2}_{2}...J^{D/2}_{N}}=\nonumber\\&& W(D,N)\epsilon^{\chi_{1}\chi_{2}..\chi_{D/2}}\nonumber\\&&\nonumber\\&& W(D,N)=\Big([\frac{(D+2)(D-2)}{8}-N(\frac{D}{2}-1)][\frac{(D+2)(D-2)}{8}-1-N(\frac{D}{2}-1)]...1\Big)\Big([N(\frac{D}{2}-1)][N(\frac{D}{2}-1)-1]...1\Big)U(\delta)\label{s11}                                                                                                                              \end{eqnarray}
                                                                                                                                                                                                                                             
                                                                                                                 where U is a function of the generalized Kronecker delta. On the other hand, $\delta S_{CGG}|_{D+1}$ has been added to the main action of supergravity to remove its anomaly. Thus, we can write:

                                                                                                                                                                                                                                          \begin{eqnarray}
                                                                                                                                                                                                                                             && \delta S_{CGG}|_{D+1}=-\delta S_{Bosonic-SUGRA}|_{D+1}=-S^{anomaly}_{Bosonic-SUGRA}|_{D+1}=0\nonumber\\&&\Rightarrow W(D,N)=\Big([\frac{(D+2)(D-2)}{8}-N(\frac{D}{2}-1)][\frac{(D+2)(D-2)}{8}-1-N(\frac{D}{2}-1)]...1\Big)\times\nonumber\\&&\Big([N(\frac{D}{2}-1)][N(\frac{D}{2}-1)-1]...1\Big)U=0 \nonumber\\&&\Rightarrow N\leq\frac{(D+2)(D-2)}{8(\frac{D}{2}-1)} \label{s12}
                                                                                                                                                                                                                                             \end{eqnarray}
                                                                                                                                                                                                                          This equation indicates that, for a ($D+1$)-dimensional space-time, the dimension of the Lie-algebra should be equal or less than a critical value. Under these conditions, the Chern-Sions gravity is free from anomalies and we do not need an extra manifold. On the other hand, as we show in (\ref{sv6}), the dimension of the algebra determines the dimension of the field-strength. This means that, for a Lie-N-algebra, field-strengths should have at most N indices. For example, for a manifold with 11 dimensions, the algebra can be of order three as predicted in recent papers \cite{b4,b9} and field-strengths may have three indices.

   \section{A Chern-Simons manifold between two 11-dimensional manifolds in an 11+3 dimensional space-time}\label{o3}
   
   In the previous Section, we have found that, for an eleven-dimensional manifold, the suitable algebra which removes the anomaly in Chern-Simons gravity is a three-dimensional Lie algebra. This means that the rank of the fields can be of order two or three. However, equation (\ref{s1}) shows that the rank of the fields may be higher than three in eleven-dimensional supergravity.  Thus, in Chern-Simons gravity theory which lives on an eleven-dimensional manifold, some extra anomalies are expected to show up. To remove them, we assume that there is another 11-dimensional manifold in the 14-dimensional space-time which interacts with the first one by exchanging Chern-Simons fields. These fields produce a Chern-Simons manifold that connects these two eleven-dimensional manifolds  ( see Fig 1.) Thus, in this model, we have two GG terms  which live on 11-dimensional manifolds (see equation (\ref{s1}))  and two CGG terms in the bulk so that each of them interacts with one of the 11-dimensional manifolds. 
    
    \begin{figure*}[thbp]
    \begin{center}
    \begin{tabular}{rl}
    \includegraphics[width=8cm]{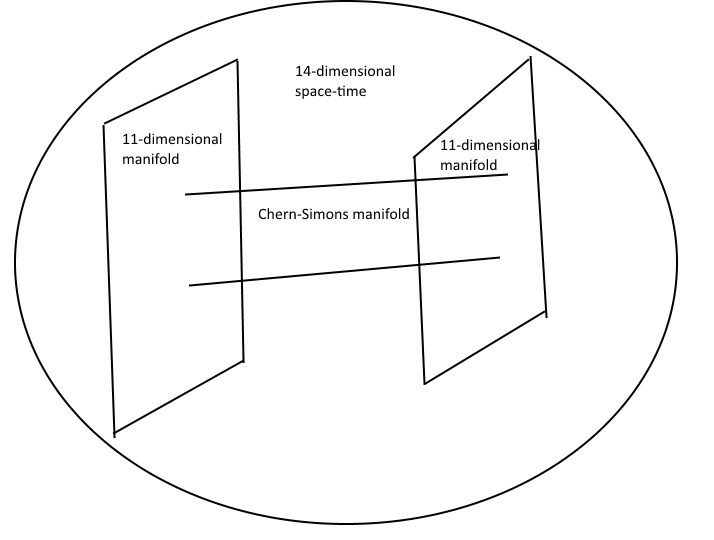}
     \end{tabular}
    \end{center}
    \caption{  Two eleven dimensional  manifolds + Chern-Simons manifold in 14-dimensional space-time. }
    \end{figure*}
    
    We can write the supergravity in 14-dimensional space-time as follows:

        \begin{eqnarray}
             && S_{SUGRA-14}= \int_{M^{N=3}}\Big(\int_{M^{11}}GG+\int_{M^{11}}\bar{C}G\bar{G}+\int_{M^{11}}C\bar{G}G+\int_{M^{11}}\bar{G}\bar{G}\Big)= \nonumber\\&& \Big(\int_{M^{14}}CGG+\int_{M^{14}}\bar{C}G\bar{G}+\int_{M^{14}}C\bar{G}G+\int_{M^{14}}\bar{C}\bar{G}\bar{G}\Big)\label{s18}
             \end{eqnarray} 
             
            In this equation, $CGG$ and $\bar{C}\bar{G}\bar{G}$ are related to the Chern-Simons gravities which live on the two eleven-dimensional manifolds and are extracting from $GG$ and $\bar{G}\bar{G}$ terms. Also, $\bar{C}G\bar{G}$ and $C\bar{G}G$ correspond to the Chern-Simon fields which are exchanged between the two manifolds in 14-dimensional space-time.  By generalizing the results of (\ref{s2}, \ref{s4} and \ref{s16}), we get:

             \begin{eqnarray}
              && G_{IJKL}\sim -(F_{IJ}F_{KL}-R_{IJ}R_{KL})+ \partial_{I}\phi\partial_{J}\phi R_{KL} ... \nonumber\\&& \delta C_{I_{i}I_{j}I_{k}} \sim - \epsilon_{I_{k}}tr(F_{I_{i}I_{j}}- R_{I_{i}I_{j}}+\partial_{I_{i}}\phi\partial_{I_{j}}\phi - \frac{1}{2}\epsilon^{I_{i}I_{j}I_{k}I_{m}} \phi R_{I_{i}I_{j}I_{k}I_{m}}+...)\nonumber\\&& \bar{G}_{IJKL}\sim -(\bar{F}_{IJ}\bar{F}_{KL}-\bar{R}_{IJ}\bar{R}_{KL})+ \partial_{I}\bar{\phi}\partial_{J}\bar{\phi} \bar{R}_{KL} ...... \nonumber\\&& \delta C_{I_{i}I_{j}I_{k}} \sim -\epsilon_{I_{k}}tr(\bar{F}_{I_{i}I_{j}}- \bar{R}_{I_{i}I_{j}}+\partial_{I_{i}}\bar{\phi}\partial_{I_{j}}\bar{\phi} - \frac{1}{2}\epsilon^{I_{i}I_{j}I_{k}I_{m}} \bar{R}_{I_{i}I_{j}I_{k}I_{m}}+...)\label{s19}
              \end{eqnarray} 
               
        Here, the $F$'s, $R$'s and $\phi$'s live on one of the supergravity manifolds as depicted in Figure 1, whereas the $\bar{F}$'s, $\bar{R}$'s and $\bar{\phi}$'s are  fields of the other supergravity manifold. To obtain their relations, we should make use of equation (\ref{s17}) and the gauge variation of the actions (\ref{s18}); in so doing, we obtain:

             \begin{eqnarray}
                  && \delta S_{SUGRA-14}= \delta \Big(\int_{M^{14}}CGG+\int_{M^{14}}\bar{C}G\bar{G}+\int_{M^{14}}C\bar{G}G+\int_{M^{14}}\bar{C}\bar{G}\bar{G}\Big)= \nonumber\\&& \int_{M^{14}}\Sigma_{n=1}^{5}\Big(tr F^{n}-tr R^{n}+tr(F^{n}R^{5-n})\Big)\nonumber\\&&+\int_{M^{14}}\Sigma_{n=1}^{5}\Sigma_{J=0}^{n}\Sigma_{m=0}^{n-J}\Big(tr \bar{F}^{J}F^{m}-tr \bar{R}^{J}R^{m}+tr(\bar{F}^{J}\bar{R}^{5-J}F^{m}R^{5-m})\nonumber\\&&+\int_{M^{14}}\Sigma_{n=1}^{5}\Sigma_{J=0}^{n}\Sigma_{m=0}^{n-J}\Big(tr \bar{F}^{m}F^{J}-tr \bar{R}^{m}R^{J}+tr(\bar{F}^{m}\bar{R}^{5-m}F^{J}R^{5-J})\nonumber\\&&+\int_{M^{14}}\Sigma_{n=1}^{5}\Big(tr \bar{F}^{n}-tr \bar{R}^{n}+tr(\bar{F}^{n}\bar{R}^{5-n})\Big)+\nonumber\\&& \int_{M^{10}+M^{N=3}}\Big(\Sigma_{n=1}^{5}(tr( F^{n}(\partial^{I_{i}}\phi\partial_{I_{j}}\phi)^{5-n}) + tr( F^{n}(\epsilon^{I_{i}I_{j}I_{k}I_{m}}  R_{I_{i}I_{j}I_{k}I_{m}}\phi)^{5-n}) \Big)+ \nonumber\\&& \int_{M^{10}+M^{N=3}}\Big(\Sigma_{n=1}^{5}(tr( \bar{F}^{n}(\partial^{I_{i}}\bar{\phi}\partial_{I_{j}}\bar{\phi})^{5-n}) + tr( \bar{F}^{n}(\epsilon^{I_{i}I_{j}I_{k}I_{m}}  \bar{R}_{I_{i}I_{j}I_{k}I_{m}}\bar{\phi})^{5-n}) \Big)+...\nonumber\\&& \approx\int_{M^{14}}\Sigma_{n=1}^{5}(F+\bar{F})^{n} - \int_{M^{14}}\Sigma_{n=1}^{5}(R+\bar{R})^{n} +\nonumber\\&& \int_{M^{14}}\Sigma_{n=1}^{5}(R\bar{F} +\bar{R}F)^{n}+..\approx \nonumber\\&& \int_{M^{14}}\Sigma_{n=1}^{5}\Big( \{ X^{I_{i}},X^{I_{j}} \}+\{ \bar{X}^{I_{i}},\bar{X}^{I_{j}} \}\Big)^{n}  =0\rightarrow   \nonumber\\&& \{ X^{I_{i}},X^{I_{j}} \}=-\{ \bar{X}^{I_{i}},\bar{X}^{I_{j}} \} \rightarrow X^{I_{i}}=i\bar{X}^{I_{i}}\rightarrow \nonumber\\&& y^{I_{i}}=i\bar{y}^{I_{i}} \quad A^{I_{i}}=i\bar{A}^{I_{i}}, \phi =i\bar{\phi} \label{s20}
                  \end{eqnarray} 
                   
       These results show that, to remove the anomaly in 14-dimensional space-time, coordinates and fields on one of the eleven-dimensional manifolds should be equal to coordinates and fields on the other manifolds in addition to one extra $i$. This implies that time- or space-like coordinates and fields on one manifold transmute into space- or time-like coordinates and fields of the another manifold. For example, the zeroth coordinate which is known as time on one manifold will transmute into a space coordinate of the other manifold. If our Universe with one time and three space coordinates is located on one of the manifolds, an anti-universe with one space and three times is located in the other manifold.
       
       Now, we shall show that, by substituting the results of equation (\ref{s20}) into the action of (\ref{s18}), the topology of the 14-dimensional manifold tends to one. This means that the world with all its matter began from a point and it may be thought of as a signature of Big Bang in our proposal. To this end, using equations (\ref{s4}, \ref{s13}, \ref{s15} and \ref{s17}), we rewrite CGG terms in terms of derivatives of scalar strings:

              \begin{eqnarray}
                   &&  \int_{M^{11}+M^{N=3}}CGG = \nonumber\\&& \int_{M^{11}+M^{N=3}}\epsilon^{I_{1}I_{2}I_{3}I_{4}I'_{1}I'_{2}I'_{3}I'_{4}I_{5}I_{6}I_{7}}\epsilon^{I_{5}I_{6}I_{7}}(\frac{\partial X^{I_{5}}}{\partial y^{I_{5}}}\frac{\partial X^{I_{6}}}{\partial y^{I_{6}}}\frac{\partial X^{I_{7}}}{\partial y^{I_{7}}}) G_{I_{1}I_{2}I_{3}I_{4}}G_{I'_{1}I'_{2}I'_{3}I'_{4}}= \nonumber\\&&\int_{M^{11}+M^{N=3}}\epsilon^{I_{1}I_{2}I_{3}I_{4}I'_{1}I'_{2}I'_{3}I'_{4}I_{5}I_{6}I_{7}}\epsilon^{I_{5}I_{6}I_{7}}(\frac{\partial X^{I_{5}}}{\partial y^{I_{5}}}\frac{\partial X^{I_{6}}}{\partial y^{I_{6}}}\frac{\partial X^{I_{7}}}{\partial y^{I_{7}}}) \times \nonumber\\&&(\epsilon^{I_{1}I_{2}I_{3}I_{4}}\frac{\partial X^{I_{1}}}{\partial y^{I_{1}}}\frac{\partial X^{I_{2}}}{\partial y^{I_{2}}}\frac{\partial X^{I_{3}}}{\partial y^{I_{3}}}\frac{\partial X^{I_{4}}}{\partial y^{I_{4}}})(\epsilon^{I'_{1}I'_{2}I'_{3}I'_{4}}\frac{\partial X^{I'_{1}}}{\partial y^{I'_{1}}}\frac{\partial X^{I'_{2}}}{\partial y^{I'_{2}}}\frac{\partial X^{I'_{3}}}{\partial y^{I'_{3}}}\frac{\partial X^{I'_{4}}}{\partial y^{I'_{4}}})=\nonumber\\&& \int_{M^{11}+M^{N=3}}\epsilon^{I_{1}I_{2}I_{3}I_{4}I'_{1}I'_{2}I'_{3}I'_{4}I_{5}I_{6}I_{7}}\epsilon^{I_{7}I_{4}}(\frac{\partial X^{I_{7}}}{\partial y^{I_{7}}})(\epsilon^{I_{4}I_{5}}\frac{\partial X^{I_{4}}}{\partial y^{I_{4}}}\frac{\partial X^{I_{5}}}{\partial y^{I_{5}}})(\epsilon^{I'_{4}I_{6}}\frac{\partial X^{I'_{4}}}{\partial y^{I_{4}}}\frac{\partial X^{I_{6}}}{\partial y^{I_{6}}}) \times \nonumber\\&&(\epsilon^{I_{1}I_{2}}\frac{\partial X^{I_{1}}}{\partial y^{I_{1}}}\frac{\partial X^{I_{2}}}{\partial y^{I_{2}}})(\epsilon^{I'_{1}I'_{2}}\frac{\partial X^{I'_{1}}}{\partial y^{I'_{1}}}\frac{\partial X^{I'_{2}}}{\partial y^{I'_{2}}})(\epsilon^{I_{3}I'_{3}}\frac{\partial X^{I_{3}}}{\partial y^{I_{3}}}\frac{\partial X^{I'_{3}}}{\partial y^{I'_{3}}})= \nonumber\\&& k\int_{M^{N=3}}\int_{M^{11}}(\delta^{11}(y)+\Sigma_{n=1}^{6}(\frac{\partial X^{I}}{\partial y^{I}})^{6-n}(F^{n}-R^{n}+...))=\nonumber\\&& k\int_{M^{N=3}}\int_{M^{11}}(\delta^{11}(y)+\Sigma_{n=1}^{6}(\frac{\partial X^{I}}{\partial y^{I}})^{6-n}( \{ X^{I_{i}},X^{I_{j}} \})^{n}) \label{s21}
                   \end{eqnarray} 
                   
          where k is a constant. There are similar results for other terms in 14-dimensional supergravity:

                \begin{eqnarray}
                     &&  \int_{M^{11}+M^{N=3}}\bar{C}\bar{G}\bar{G} =  \nonumber\\&& k\int_{M^{N=3}}\int_{M^{11}}(\delta^{11}(y)+\Sigma_{n=1}^{6}(\frac{\partial \bar{X}^{I}}{\partial \bar{y}^{I}})^{6-n}(\bar{F}^{n}-\bar{R}^{n}+..))= \nonumber\\&& k\int_{M^{N=3}}\int_{M^{11}}(\delta^{11}(y)+\Sigma_{n=1}^{6}(\frac{\partial \bar{X}^{I}}{\partial \bar{y}^{I}})^{6-n}(\{ \bar{X}^{I_{i}},\bar{X}^{I_{j}} \}))^{n} \label{s22}
                     \end{eqnarray}

                    \begin{eqnarray}
                         &&  \int_{M^{11}+M^{N=3}}\bar{C}G\bar{G} =  \nonumber\\&& k\int_{M^{N=3}}\int_{M^{11}}(\delta^{11}(y)+\Sigma_{n=1}^{6}\Sigma_{m=0}^{n}(\frac{\partial \bar{X}^{I}}{\partial \bar{y}^{I}})^{6-n}(\bar{F}^{m}-\bar{R}^{m}+...)(F^{n-m}-R^{n-m}+...))=\nonumber\\&& k\int_{M^{N=3}}\int_{M^{11}}(\delta^{11}(y)+\Sigma_{n=1}^{6}\Sigma_{m=0}^{n}(\frac{\partial \bar{X}^{I}}{\partial \bar{y}^{I}})^{6-n}(\{ \bar{X}^{I_{i}},\bar{X}^{I_{j}} \})^{n-m}( \{ X^{I_{i}},X^{I_{j}} \})^{m}) \label{s23}
                         \end{eqnarray}

                                      \begin{eqnarray}
                                           &&  \int_{M^{11}+M^{N=3}}C\bar{G}G =   \nonumber\\&& k\int_{M^{N=3}}\int_{M^{11}}(\delta^{11}(y)+\Sigma_{n=1}^{6}\Sigma_{m=0}^{n}(\frac{\partial \bar{X}^{I}}{\partial \bar{y}^{I}})^{6-n}(\bar{F}^{n-m}-\bar{R}^{n-m}+...)(F^{m}-R^{m}+...))=\nonumber\\&& k\int_{M^{N=3}}\int_{M^{11}}(\delta^{11}(y)+\Sigma_{n=1}^{6}\Sigma_{m=0}^{n}(\frac{\partial \bar{X}^{I}}{\partial \bar{y}^{I}})^{6-n}(\{ \bar{X}^{I_{i}},\bar{X}^{I_{j}} \})^{m}( \{ X^{I_{i}},X^{I_{j}} \})^{n-m}) \label{s24}
                                           \end{eqnarray}
                                            
        Using the results in equations (\ref{s20}) and substituting equations (\ref{s21},\ref{s22},\ref{s23}and \ref{s24}) in equation (\ref{s20}), we obtain:

              \begin{eqnarray}
                   && (\frac{\partial X^{I}}{\partial y^{I}})=(\frac{\partial \bar{X}^{I}}{\partial \bar{y}^{I}})\rightarrow \nonumber\\&& S_{SUGRA-14}= k\int_{M^{N=3}}\int_{M^{11}}\Big(\delta^{11}(y)+\Sigma_{n=1}^{6}(\frac{\partial X^{I}}{\partial y^{I}})^{6-n}\Big( \{ X^{I_{i}},X^{I_{j}} \}+\{ \bar{X}^{I_{i}},\bar{X}^{I_{j}} \}\Big)^{n}\Big)\label{s25}
                   \end{eqnarray} 
                   
        On the other hand, results in equation (\ref{s20}) show that ($\{ X^{I_{i}},X^{I_{j}} \}=-\{ \bar{X}^{I_{i}},\bar{X}^{I_{j}} \}$).  Thus, we can conclude that the action given above tends to an action on the three-dimensional manifold.
        
        \begin{eqnarray}
                    && \{ X^{I_{i}},X^{I_{j}} \}=-\{ \bar{X}^{I_{i}},\bar{X}^{I_{j}} \} \nonumber\\&& S_{SUGRA-14}= k\int_{M^{N=3}}\int_{M^{11}}\Big(\delta^{11}(y)\Big)=\nonumber\\&& k\int_{M^{N=3}}\rightarrow k \frac{4\pi R^{3}}{3}\label{s25}
                    \end{eqnarray}
                     
        where R is the radius of manifold. By redefining scalar strings, we are able to show that the action of supergravity shrinks to one.

        \begin{eqnarray}
                    &&  X\rightarrow (k \frac{4\pi R^{3}}{3})^{-1/11}X \nonumber\\&& \bar{X}\rightarrow (k \frac{4\pi R^{3}}{3})^{-1/11}\bar{X}\Rightarrow \nonumber\\&& S_{SUGRA-14}= (k \frac{4\pi R^{3}}{3})^{-1}k\int_{M^{N=3}}\rightarrow (k \frac{4\pi R^{3}}{3})^{-1} k \frac{4\pi R^{3}}{3}=1\label{s26}
                    \end{eqnarray}
                    
          \begin{figure*}[thbp]
              \begin{center}
              \begin{tabular}{rl}
              \includegraphics[width=8cm]{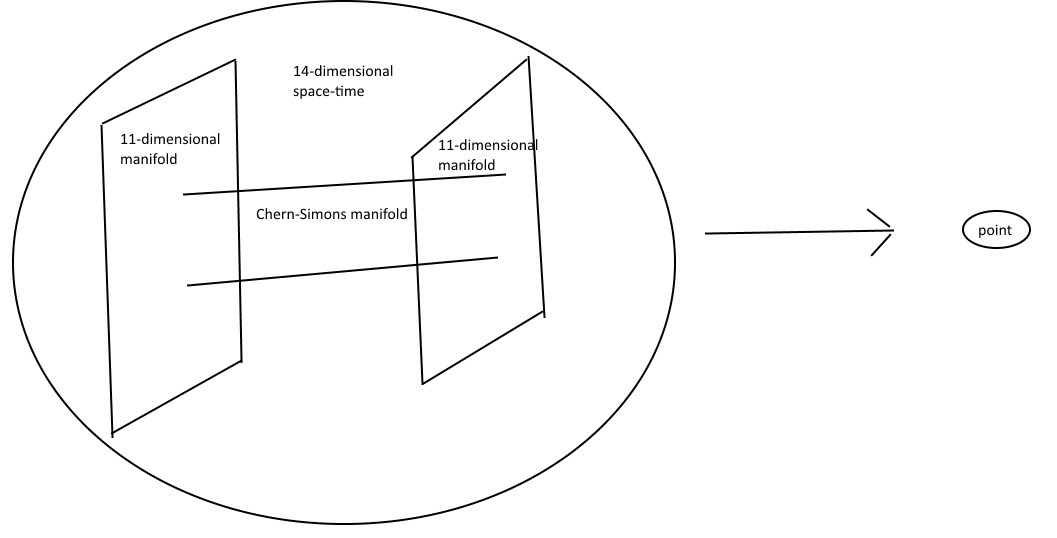}
               \end{tabular}
              \end{center}
              \caption{  14-dimensional manifold shrinks to one point. }
              \end{figure*}

        This equation yields results that deserves our comments. In fact, two eleven-dimensional manifolds and the bulk interact with each other via different types of C, G and Chern-Simons-fields. When we sum over supergravities that live on these manifolds and consider fields in the space between them, we get supergravity in 14-dimensional space. By canceling the anomaly in this new supergravity, we can obtain the relations between fields. By substituting these relations into the action of the 14-dimensional supergravity, we simply obtain one. This means that the 14-dimensional manifold with all its matter content can be topologically shrunk to one point (See Fig.2.). In fact, the system of the world began from this point and then expand and construct a 14-dimensional world similar to what happens in a Big Bang theory.

\section{Summary and Final Considerations }\label{o4}
In this paper, we have shown that the Chern-Simons terms of modified gravity may be understood as due to the interaction between two 11-dimensional manifolds in an 11+3-dimensional space-time, where 3 is the dimension of a Lie-type-algebra. We also argue that there is a direct relation between the dimension of the algebra and the dimension of the manifold. For example, for 11-dimensional world, the dimension of the Lie-agebra is three. If the rank of the fields which live in one manifold becomes larger than the rank of the algebra, there emerges an anomaly. This anomaly is produced as an effect of connecting fields in the bulk to fields which live in the manifold. To cancel this anomaly, we need to introduce another 11-dimensional manifold in the 11+3-dimensional space-time which interacts with the initial manifold by exchaning Chern-Simons terms. These Chern-Simons terms produce an extra manifold. If we sum over the topology of the 11-dimensional manifolds and the topology of the Chern-Simons manifold, we can show that the total topology shrinks to one, which is consistent with predictions of the Big Bang theory.

To conclude, we would like to point out that that all our treatment has been restricted to the purely bosonic sector of 11-dimensional supergravity, whose on-shell multiplet, besides the metric tensor and the 3-form gauge potential, includes the gravitino field. The latter has not been considered here; we have restricted ourselves to the bosonic fields. However, it would be a further task to inspect how the inclusion of the gravitino would affect our approach, possibly changing the dimension of the Lie algebra to be added to the 11-dimensional manifold. By including the fermion, it is no longer ensured that the local supersymmetry of the 11+3-dimensional supergravity action remains valid. The change in the dimension of the Lie algebra could, in turn, give rise to new terms, so that the Chern-Simons modified gravity would be further extended as a result of including the gravitino. We intend to pursue an investigation on this issue and to report on it elsewhere.

\section*{Acknowledgements}
\noindent The work of Alireza Sepehri has been supported
financially by the Research Institute for Astronomy and Astrophysics
of Maragha (RIAAM),Iran under the Research Project No.1/4165-14.


\begin{thebibliography}{99}
\bibitem{tt1}R. Jackiw and S. Y. Pi, Phys. Rev. D 68 (2003) 104012.
\bibitem{tt2}B. Pereira-Dias, C. A. Hernaski, J. A. Helayël-Neto, Phys.Rev.D83:084011,2011.
\bibitem{tt3}Kohkichi Konno, Toyoki Matsuyama, Satoshi Tanda, Phys.Rev.D76:024009,2007.
\bibitem{tt4}David Guarrera, A. J. Hariton, Phys.Rev.D76:044011,2007.
\bibitem{tt5}K.K. Nandi, I.R. Kizirgulov, O.V. Mikolaychuk, N.P. Mikolaychuk, A.A. Potapov, Phys.Rev.D79:083006,2009.
\bibitem{tt6}P. J. Porfirio, J. B. Fonseca-Neto, J. R. Nascimento, A. Yu. Petrov, J. Ricardo, A. F. Santos,  Phys. Rev. D 94, 044044 (2016).
\bibitem{tt7}N. Yunes, F. Pretorius, Phys. Rep. 480, 1 (2009), arXiv: 0907.2562.\\
 Songbai Chen, Jiliang Jing, Class. Quant Grav.27:225006, 2010.
\bibitem{tt8}Carlos F. Sopuerta, Nicolas Yunes, Phys.Rev.D80:064006,2009.
\bibitem{b1}Petr Horava, Edward Witten, Nucl. Phys. B460 (1996) 506, hep-th/9510209.
\bibitem{b2}Petr Horava, Edward Witten, Nucl.Phys.B475:94-114,1996.
\bibitem{b4}J. Bagger and N. Lambert, Gauge Symmetry and Supersymmetry of Multiple M2-Branes,
Phys. Rev. D 77, 065008 (2008) [arXiv:0711.0955 [hep-th]].

\bibitem{b9}Pei-Ming Ho, Yutaka Matsuo, JHEP 0806:105,2008.
\bibitem{A1}Alireza Sepehri, Anomaly cancellation in D+N dimensional supergravity, submitted to journal, in proceeding;\\ Alireza Sepehri, Richard Pincak, accepted in Modern Physics Letters A, arXiv:1610.09277 [physics.gen-ph]; \\ Alireza Sepehri, Richard Pincak,  arXiv:1610.09257 [physics.gen-ph].

\end{thebibliography}
\end{document}